\documentclass[8pt,preprint]{article}
\usepackage{graphicx}
\usepackage{amssymb}
\def\beq{\begin{eqnarray}}
\def\eeq{\end{eqnarray}}

\usepackage{amsfonts}

\usepackage[utf8]{inputenc}
\usepackage[utf8]{inputenc}
\usepackage[T1]{fontenc}
\usepackage{subfigure}
\usepackage[usenames,dvipsnames]{xcolor}
\usepackage{tikz}
\usepackage{pgf}
\usepackage{booktabs}
\usepackage{slashed}
\usepackage{rotating}
\usepackage{colortbl}
\usepackage{verbatim}
\usepackage{url}
\usepackage[retainorgcmds]{IEEEtrantools}
\usepackage{amsmath, amsthm, amsfonts,amssymb}
\usepackage{graphicx}
\usepackage{appendix}
\usepackage{braket}
\usepackage{feynmp}
\usepackage{bbm}

\begin{document}


\title{Weakly (and not so weakly) bound states of a relativistic particle in one dimension}
\author{Paolo Amore \\
\small Facultad de Ciencias, CUICBAS, Universidad de Colima,\\
\small Bernal D\'{i}az del Castillo 340, Colima, Colima, Mexico \\
\small paolo.amore@gmail.com \\
Francisco M. Fern\'andez \\
\small INIFTA (CONICET),  Divisi\'on Qu\'imica Te\'orica, \\
\small Blvd. 113 S/N, Sucursal 4, Casilla de Correo 16, 1900 La Plata, Argentina \\
\small framfer@gmail.com \\
Enrique Jimenez \\
\small Facultad de Ciencias, CUICBAS, Universidad de Colima,\\
\small Bernal D\'{i}az del Castillo 340, Colima, Colima, Mexico \\
\small physieira@gmail.com}

\maketitle

\begin{abstract}  
We present the first exact calculation of the  energy of the bound state of a one dimensional Dirac massive particle  in weak short-range 
arbitrary potentials, using perturbation theory to fourth order (the analogous result for two dimensional systems with confinement along 
one direction and arbitrary mass is also calculated to second order). We show that the non--perturbative extension obtained using
Pad\'e approximants can provide remarkably good approximations even for deep wells, in certain range of physical parameters. 
As an example, we discuss the case of two gaussian wells, comparing numerical and analytical results, predicted by our formulas. 
\end{abstract}

Almost 90 years have passed since Dirac established his famous equation, successfully combining 
Quantum Mechanics and Special Relativity, the two physical theories that completely changed 
our understanding of Nature at the beginning of the previous century. The importance of the Dirac 
equation can hardly be overstated: it predicts the existence of antimatter (discovered by Anderson in 1932), 
it explains the spin of the electron, recovering Pauli's theory in the low energy limit, and it also describes
correctly the observed spectrum of the hydrogen atom, all at once. Another consequence of the Dirac equation,
the \textit{Zitterbewegung} ({\sl trembling motion}) of the electron,  has not been experimentally observed,
although recently it has been simulated on physical systems composed of atoms which mimic the behavior of 
a free relativistic particle~\cite{Gerritsma2010, Dreisow2010}. In recent years, the Dirac equation has also 
been used to describe the low energy spectrum of graphene, with either massless~\cite{Ando2006} or 
massive~\cite{Benfatto2008}  excitations.

It is interesting to observe that even from the point of view of the theory, there are consequences of the Dirac
equation that still need to be explored; our attention in the present paper is devoted to the study of the
behavior of weakly bound relativistic states in one and two dimensional systems. 
The non-relativistic counterpart of this problem, has been settled long time ago in a seminal paper by 
Simon~\cite{Simon1976TheBound}, where the conditions for the existence of this bound state have been given and
the analyticity (non-analyticity) of the energy in one (two) dimension has been established. 

For the relativistic case, the conditions under which a Dirac particle is trapped in a one-dimensional potential 
have  been identified in ref.~\cite{coutinho1987conditions}; more recently Cuenin and Siegel~\cite{Cuenin2017} 
have studied the weakly coupling eigenvalue asymptotics for the bound state of the one dimensional Dirac operator, 
perturbed by a matrix-valued and non-symmetric potential. 

For the case of a non-relativistic particle in a one dimensional short-range potential, a formula 
for the energy of the bound state has been derived up to sixth order: Simon~\cite{Simon1976TheBound} 
reports an unpublished result obtained by Abarbanel, Callan and Goldberger~\cite{Abarbanel}, which is 
exact to third order in the parameter  controlling the strength of the potential, whereas higher order 
corrections (up to order six) have been derived later~\cite{Patil80,Gat93,Amore2017Perturbation} using 
different techniques. Interestingly, a similar analysis for the relativistic case is still lacking 
and this constitutes the main goal of the present paper.

The approach that we will follow in this paper has been originally proposed by Gat and Rosenstein~\cite{Gat93}, 
and applied to the non-relativistic version of the present problem (to third order in the perturbation 
parameter) and to a $(1+1)$ dimensional QFT; in a recent work by two of the present authors, 
ref.~\cite{Amore2017Perturbation}, the method has been applied to calculate the energy of the bound state 
of an arbitrary shallow short range potential to sixth order. 

We will first briefly describe how the method works for the non-relativistic problem and then discuss how 
it can be extended to its relativistic counterpart. 

Let $\hat{H}$ be the hamiltonian of the problem 
\begin{eqnarray}
\hat{H}(\lambda) =  - \frac{d^2}{dx^2} + \lambda V(x) 
\end{eqnarray}
where $V(x) < 0$ for $x \in (-\infty, \infty)$ and $\lim_{|x| \rightarrow \infty } V(x) = 0$. Here $\lambda>0$
is a parameter that controls the strength of the potential well. As noticed in \cite{Gat93}, one cannot
use  $\hat{H}(0)$ as the unperturbed hamiltonian, since, for $\lambda >0$ the spectrum of $\hat{H}$ contains
(at least) one bound state, whereas the spectrum of $\hat{H}(0)$ is continuous. 

Instead we use as unperturbed Hamiltonian the operator
\begin{eqnarray}
\hat{H}_0 \equiv  - \frac{d^2}{dx^2}  - 2\beta \delta(x) \ \ \ , \ \ \ \beta>0 \ \  \ .
\end{eqnarray}

$\hat{H}_0$ has just one bound state with energy $\epsilon_0 = -\beta^2$ and a continuum 
of states, for $\epsilon > 0$. As a result, the Schr\"odinger equation
\begin{eqnarray}
\left[ \hat{H}_0 + \lambda V(x)  \right] \psi(x) = E \psi(x)
\end{eqnarray}
can now be studied perturbatively in $\lambda$, working with a finite $\beta$ and 
assuming $E = \sum_{n=0}^\infty \lambda^n \epsilon_n$ and $\psi(x) = \sum_{n=0}^\infty \lambda^n \phi_n(x)$.
The infrared divergencies, which would spoil the perturbative expansion when $H(0)$ is used, 
manifest, at a given order, as inverse powers of $\beta$, and cancel out exactly, rendering each order 
perfectly finite.

Contrary to the approach followed in \cite{Gat93, Amore2017Perturbation}, where the standard Rayleigh-Schr\"odinger
approach involving matrix elements was applied, here we obtain a perturbative solution of the Schr\"odinger equation 
in terms of the appropriate Green's funcions.

To lowest order in $\lambda$ one has the eigenvalue equation
\begin{IEEEeqnarray}{rl}
    \left(-\frac{d^{2}}{dx^{2}} \, - \, \beta\,\delta(x) \right)\phi_{0}(x)  \, = \, \epsilon_{0}\phi_{0}(x)
    \label{Schrodinger Modi  lambda zero}
\end{IEEEeqnarray}

In this case the eigenvalue and eigenfunction are  $ \epsilon_{0}=-\beta^{2} $ and 
$\phi_{0}(x)  \, = \, \sqrt{\beta} e^{-\beta \vert x \vert}$ respectively.

To higher orders one obtains the equations
\begin{eqnarray}
\mathcal{D} \phi_n(x) = - V(x) \phi_{n-1}(x) +  \sum_{k=1}^n  \epsilon_k \phi_{n-k}(x) \equiv
\mathcal{S}_n(x) 
\end{eqnarray}
with $\mathcal{D} \equiv \left(  - \frac{d^2}{dx^2} - 2 \beta \delta(x) +\beta^2 \right)$.
To deal with them one needs to consider the Green's function $G(x,y)$ defined
by
\begin{IEEEeqnarray}{rl}
   \mathcal{D} G(x,y) \, = \, \delta(x-y)
    \label{Green Function}
\end{IEEEeqnarray}
and write the solution of order $ n $ as $\phi_{n}(x)  \, = \, \int G(x,y)\mathcal{S}_{n}(y)dy$.
The exact form of this and higher orders Green's functions can be found
in ref. \cite{Amore2017Perturbation}.
This equation needs to be complemented by the condition
\begin{IEEEeqnarray}{rl}
            \int \mathcal{S}_{n}(x)   \phi_{0}(x) dx  \, = \,0\, \, ; \, \, n\geq 1\ .
    \label{secular terms}
\end{IEEEeqnarray}
which removes the "secular terms" in the expansion. 
Equation (\ref{secular terms}) only gives the energy and the wave function at a given order.

This approach has the advantage of avoiding the appearance of infinite series and it allows one to consider
more general eigenvalue equations, as in the case of a relativistic particle.

Let us now discuss the case of a relativistic particle in one or two dimensions, obeying the  Dirac equation 
$ \hat{H}\psi=E(\lambda)\psi $, where
\begin{IEEEeqnarray}{rl}
          \hat{H}  &\, = \,  - i\, \sigma\cdot\nabla \, + \, \sigma_{3}\, m   \, + \, {\lambda}W(x)\, 
    \label{Dirac Eq.}
\end{IEEEeqnarray}
and $\psi = (\psi_1 \ \psi_2)$ is a spinor ($\sigma_i$ are the usual Pauli matrices).

Here $ \sigma\cdot\nabla = \sigma_1 \partial_x $ for the one dimensional case and
$\sigma\cdot\nabla = \sigma_1 \partial_x + \sigma_2 \partial_y $ for the two-dimensional one.

The potential, which depends only on $x$, is given by 
\begin{eqnarray}
W(x) = \frac{1}{2} \left[ \sigma_{3} \left( \,V(x)+U(x) \right) + \mathbbm{1} \left(  \,V(x)-U(x)\right)\right] \ ,
\end{eqnarray} 
where $(V(x)+U(x))/2$ and $(V(x)-U(x))/2$ are a vector and a scalar potential respectively.

Equations of the form of (\ref{Dirac Eq.}) have been studied previously, in particular for the case of 
point-like interactions in one dimension~\cite{Dominguez1989}  and for graphene and graphite systems,  
subject to piecewise-constant potentials~\cite{Pereira2006,Yampolskii2008}. 

We can work in one or two dimensions in an unified framework by using the ansatz $ \exp[iq y] \psi(x)$ 
(the one dimensional case is recovered for $q=0$) and write explicitly the Dirac equation in terms of its components
\begin{eqnarray}
  (-E+m+\lambda V) \psi_{1} - i \left(q + \partial_{x}\right)\psi_{2} &=& 0 \nonumber\\
- (E+m+\lambda  U) \psi_{2} + i \left(q - \partial_{x}\right)\psi_{1} &=& 0
\label{components dirac}
\end{eqnarray}

Using the second equation we can express $\psi_2$ in terms of $\psi_1$ and then use it inside the first equation
to obtain a second order differential equation for $\psi_1$ alone:
\begin{eqnarray}
-\psi_{1}{''}(x)  &+&  \frac{\lambda  U'(x) \psi_1'(x)}{E+m+\lambda  U(x)}
+ \left( - \frac{\lambda  q U'(x)}{{E}+m+\lambda  U(x)} +\lambda  (m-E) U(x) \right. \nonumber \\
&+& \left.  \lambda  (E+m) V(x)+\lambda^2 U(x) V(x) \right) \psi_1(x) \nonumber \\
&=&\left( E^{2}-k^{2}(q)\right)\psi_{1}(x)  ,
    \label{Decoupled Dirac Eq} 
\end{eqnarray}
with  $k(q) \equiv \sqrt{q^2+m^2}$. For the special case $U(x) = 0$ this equation takes a simpler form
of a Schr\"odinger--like equation, with an energy dependent potential,
as already pointed out by Coutinho and Nogami~\cite{coutinho1987conditions}.

Eq. \eqref{Decoupled Dirac Eq} is now in the appropriate form to be attacked using the approach that we have previously 
described for the non-relativistic case. We cast this equation in a compact form, formally similar to the nonrelativistic case, as
\begin{IEEEeqnarray}{rl}
              \mathcal{D}\,\psi_{1} &\, = \, \mathcal{V}\, \psi_{1} \ , 
    \label{D and S operators}
\end{IEEEeqnarray}
where $\mathcal{V}(x)$ can be read off the equation (\ref{Decoupled Dirac Eq}).

After expressing both the energy and wave function as power series in $\lambda$,
\begin{IEEEeqnarray}{rl}
       E= \sqrt{k^{2}- \Delta}  , \quad      \Delta & \, = \, \sum^{\infty}_{n=0}\delta_{n}\,\lambda^{n} ,\quad
\psi_{1} \, = \, \sum^{\infty}_{n=0}\phi_{n}\,\lambda^{n}  \nonumber
\end{IEEEeqnarray}
and substituting into eq.~(\ref{D and S operators}), we obtain a infinite tower of second order differential equations, 
corresponding to different orders in $\lambda$,which can be solved starting from the lowest order.  
This situation is completely analogous to the non-relativistic case, although now $\mathcal{V}$ 
in eq.(\ref{D and S operators}) is  an operator and it is non-linear in $\lambda$. The main consequence of this fact is the 
rapid proliferation of terms contributing at a given perturbative order, as the order is increased.

Applying the method of Gat and Rosenstein to this equation, we have obtained the perturbative expression for 
the energy of the fundamental mode to fourth order in $\lambda$ for the one-dimensional problem and to second
order in $\lambda$ for the two-dimensional model.

Since the solutions for the one-dimensional case can be recovered from the corresponding two-dimensional expressions
setting $q=0$, we first assume $|q| \geq 0$, and report the coefficients  of $ \Delta $ up to second order
\begin{IEEEeqnarray}{rl} 
\delta_{0} &\, = \, \beta^{2}\ ,\nonumber \\
\delta_{1} & \, = \, -2 \beta \,\mathcal{F}(k)   \, + \, \mathcal{O}(\beta^{2})\ ,  \nonumber \\
\delta_{2} &  \, = \, \mathcal{F} (k) ^{2}  \, + \,\mathcal{O}(\beta)\ ,   
\label{delta 1 to delta 3}
\end{IEEEeqnarray}
where 
\begin{IEEEeqnarray}{rl}
            \mathcal{F}(k)   \, = \,   \tfrac{1}{2}\int dx\, \left( \left( m+k\right)V \, + \,  \left( m-k\right)U\right) \ .
    \label{F1 kappa}
\end{IEEEeqnarray}

The coefficients $\delta_i$ obtained in the calculation, with $i=1,2,\dots$, are analytic 
functions of $\beta$, $\delta_i = \sum_{n=0}^\infty \delta_i^{(n)} \beta^n$. Although the Green's functions contain singular 
contributions at $\beta =0$, the energy obtained in perturbation theory does not contain infrared divergences 
for $\beta\rightarrow 0^+$, due to exact cancellations, as for the non-relativistic case. 

For the one-dimensional case ($ q=0 $), we have computed the energy corrections up to order four.  Terms up to second  order 
are obtained by making  $ \kappa=m $ in formula \eqref{delta 1 to delta 3}, while  $ \delta_{3} $ and  $ \delta_{4} $   are given by
\begin{IEEEeqnarray}{rl} 
\delta_{3}   & \, = \, 2 m^3\,\mathcal{F}_{1} \mathcal{F}_{2,1}  \, + \,\mathcal{O}(\beta)\ ,     \nonumber \\
\delta_{4}   & \, = \, {m^4} \,\eta_{4} \, \, - \, {  m^2}\kappa_{4}   \, + \,\mathcal{O}(\beta)\ ,      
\label{delta 3 and delta 4}
\end{IEEEeqnarray}
where
\begin{IEEEeqnarray}{rl}
\eta_{4} &\,  \, = \,  \, \left( \mathcal{F}_{1} \right) ^{2} \mathcal{F}_{2,2}\, + \, 2 \,\mathcal{F}_{1}  \mathcal{F}_{3,1}\, 
                 + \,  \left( \mathcal{F}_{2,1}\right) ^{2} \ , \nonumber \\
\kappa_{4} & \, = \, (1/2) \left(\mathcal{F}_{1}     \mathcal{F}_{3,2}\, + \, \left( \mathcal{F}_{1} \right)  ^4\right)  \ .
    \label{delta4}
\end{IEEEeqnarray}

$ \mathcal{F}_{1},\dots, \mathcal{F}_{3,2} $ are functionals of $ V $ and $ U $ given by 
\begin{eqnarray}
\mathcal{F}_{1}   &=& \mathcal{F}(m)/m =  \int dx\, V(x) \ , \nonumber\\
\mathcal{F}_{2,1} &=& \int\int \,  dx dy\, V(y)\left| x-y\right|  V(x) \ , \nonumber \\
\mathcal{F}_{2,2} &=& \int\int \,  dx dy\, V(y)\left( x-y\right)^{2}  V(x)  \ ,\nonumber \\
\mathcal{F}_{3,1} &=& \int\int \int \,  dx dy dz\, \left| x-y\right|  \left| y-z\right| V(x)V(y)  V(z)\ , \nonumber \\
\mathcal{F}_{3,2} &=& \int\int\int dx dy dz \frac{\left\vert x-y\right\vert \left\vert x-z\right\vert}{(x-y)(x-z)}U(x) V(z) V(y) \ .    \nonumber 
\end{eqnarray}

The energy of the bound state in one dimension reads
\begin{IEEEeqnarray}{rl}
E^{(1)}(\lambda)  \, = \,  m+ \tilde{E}(\lambda) \, + \, \lambda^{4} \,\delta E\, + \, \mathcal{O}(\lambda^{5}) \ ,
\label{energy bound}
\end{IEEEeqnarray}
where
\begin{eqnarray}
\tilde{E} &=& -\frac{m\lambda^2}{2} \mathcal{F}_{1}^2   - m^2 \lambda^3 \mathcal{F}_{1} \mathcal{F}_{2,1} 
- \frac{m^3 \lambda^4}{2} \eta_4
\end{eqnarray}
is the non-relativistic formula previously obtained to fourth order working with the Schr\"odinger equation
and $\delta E$ is the leading relativistic correction which appears to fourth order
\begin{eqnarray}
\delta E= \frac{m}{2} \left( \kappa_{4} - \frac{1}{4} \mathcal{F}_{1}^4\right)
\end{eqnarray}

Note that, while $\tilde{E}(\lambda)$ is a functional of $V$ only, $\delta E$ is a functional both of $V$ and $U$.

In two dimensions, for quasi-bound states of the form $ \psi(x,y)  =\exp(iq)\psi(x) $, the energy is given by 
\begin{IEEEeqnarray}{rl}
E^{(2)}(\lambda)  = k  - ({\lambda^{2}}/{2 k })\,\mathcal{F}(k) ^{2}  \, + \, \mathcal{O}(\lambda^{3})\ .
\label{energy quasi-bound}
\end{IEEEeqnarray}

For the case of a relativistic one-dimensional square well, discussed by Greiner  \cite{Greiner1990} in detail, Eq.~(\ref{energy bound}) 
reproduces the exact results up to fourth order.
As a further test of our perturbation expressions we also consider the simple
case in which $V(x)=-(1+\gamma )\delta (x)$, $U(x)=-(1-\gamma )\delta (x)$
and $q=0$. In order to avoid the possible discontinuity of both functions 
$\psi _{1,2}(x)$ at $x=0$ we set $\gamma =1$. In this case $\psi _{1}(x)$ is 
continuous at $x=0$ and a straightforward calculation shows that
\begin{equation}
E^{(1)}=m\frac{1-\lambda ^{2}}{1+\lambda ^{2}}=m-2m\lambda ^{2}\left( 1-\lambda
^{2}+\lambda ^{4}+\ldots \right) ,  \label{eq:E_series}
\end{equation}%
and $\psi (x)=\sqrt{\beta }e^{-\beta |x|}$, with $\beta =\sqrt{m^{2}-E^{2}}=\frac{2m\lambda }{1+\lambda ^{2}}$.

Note that $0<\beta \leq \beta (\lambda =1)=m$. Present perturbation theory
yields the first three terms of the series~(\ref{eq:E_series}) exactly.

In the perturbative region, $0 < \lambda \ll 1$, the relativistic correction $\delta E$, provides in general a tiny correction
to the corresponding non-relativistic expression, $\tilde{E}$, implying that the weakly bound electron is essentially 
non-relativistic. This hierarchy can however be modified already at moderate values of $\lambda$. 
In this case, the energy of the bound state cannot lower indefinitely as the well becomes deeper and deeper, as in the non--relativistic case, since it is trapped between two 
continua, the continuum of positive energy states, for $E \geq m$, and the continuum of negative energy states, for $E \leq -m$.
This behavior can be captured using a diagonal Pade approximant, which tends to a constant for $\lambda \rightarrow \infty$:
\begin{eqnarray}
E^{(1)}_{\rm Pade} &=& m  + \frac{m^2\lambda^2 \mathcal{F}_1^4}{-2 m \mathcal{F}_1^2
+ 4 m^2 \lambda  \mathcal{F}_1 \mathcal{F}_{2,1}
+ 2 \lambda^2 \left(-2 \delta E +m^3 \left(\eta_4 -4 \mathcal{F}_{2,1}^2\right)\right)}
\label{pade}
\end{eqnarray}

This formula provides a completely analytical expression for the energy of the relativistic bound state which can 
be used for larger values of $\lambda$; the non relativistic case can either be obtained setting $\delta E\rightarrow 0$ 
in this expression, or using the simpler $[2,1]$ Pade approximant, which is linear as $\lambda \rightarrow \infty$:
\begin{eqnarray}
E^{(1)}_{\rm Pade-nr} &=& - \frac{m \lambda^2 \mathcal{F}_1^3}{4 m \lambda \mathcal{F}_{2,1} -2 \mathcal{F}_1}
\label{Pade-nr}
\end{eqnarray}

One way to assess the region of applicability of Eq.~(\ref{pade}) is by identifying the region in parameter space where
\begin{eqnarray}
\delta E > \frac{m^3}{2} \left( \eta_4 - 3 \mathcal{F}_{2,1}^2 \right)
\label{ineq0}
\end{eqnarray}
is fulfilled. When this condition is met, the denominator of $E^{(1)}_{\rm Pade}$ has no real pole and consequently
the resummation is more accurate. 

As an example, we consider the gaussian wells $V(x) = - (1+\gamma) e^{-\alpha x^2}$ and  
$U(x) = - (1-\gamma) e^{-\alpha x^2}$, where $-1 \leq \gamma \leq 1$ is a parameter which controls 
the depths of $V(x)$ and $U(x)$~\footnote{Note that the case
$|\gamma|>1$  can be reduced to the present case by means of a  redefinition of $\lambda$.}.

In this case the inequality (\ref{ineq0}) reads
\begin{eqnarray}
\pi  \alpha  (\gamma +5) > 8 \left(-6+3 \sqrt{3}+2 \pi \right) (\gamma +1) m^2
\label{ineq}
\end{eqnarray}

\begin{figure}[h]
\centering
\includegraphics[width=0.6\textwidth]{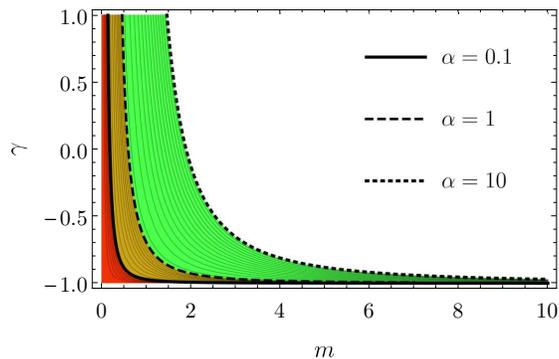}
\caption{Parameter space for the inequality (\ref{ineq}) for three different values of $\alpha$. (color online)}
\label{fig:ineq}
\end{figure}

The region in parameter space where the inequality is fulfilled is displayed in Fig.~\ref{fig:ineq}, for three
values of $\alpha$. Notably the Pad\'e has always real poles when $\delta E$ is set to zero.

\begin{figure}[h]
\centering
\includegraphics[width=0.6\textwidth]{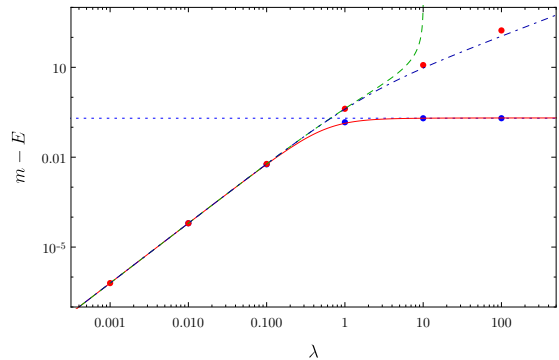} 
\caption{Energy of the relativistic and non-relativistic bound states (blue and red dots respectively) compared with the
relativistic and non-relativistic Pad\'e approximants (solid, dashed and dot-dashed lines respectively).
Here $m=0.1$, $\alpha=1$ and $\gamma=1$.
The horizontal lines are the limit values $m-E = m (1+\gamma)$. (color online)}
\label{fig:en}
\end{figure}

In Fig.~\ref{fig:en} we plot the quantity $m-E$ for the case $m=0.1$ and $\alpha=1$, as a function of $\lambda$, for
$\gamma=1$ (the plot for $\gamma=0$, not reported here is quite similar). The blue points represent the numerical values obtained
applying the shooting method directly to Eq.~(\ref{Decoupled Dirac Eq}); the red points are obtained solving the
corresponding non-relativistic Schr\"odinger equation. These values are compared with the relativistic Pad\'e of Eq.~(\ref{pade}) (solid line),
the non-relativistic $[2,2]$ Pad\'e obtained setting $\delta E=0$ (dashed line) and the non--relativistic Pad\'e 
of Eq.~(\ref{Pade-nr}) (dot-dashed line).
The horizontal lines correspond to the limit values $m (1+\gamma)$. 
While $\delta E$ provides a tiny contribution at small $\lambda$, it plays an essential role at larger values 
of $\lambda$.

\begin{figure}[h]
\centering
\includegraphics[width=0.6\textwidth]{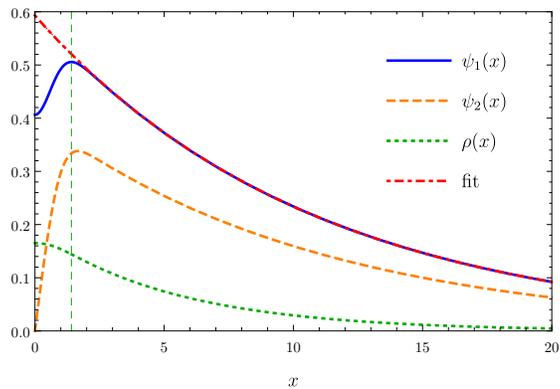} 
\caption{Normalized upper and lower components $\psi_{1,2}(x)$ and probability density 
of the Dirac spinor for  $m=0.1$, $\alpha=1$, $\gamma=0$ and $\lambda=1$, 
obtained using the shooting method. The vertical line is the location of the singularity $E+m+\lambda U(x)$, where $\psi_1'(x)=0$. 
The dashed line is the fit of the numerical results between $x=5$ and $x=50$, 
$\psi_1^{(fit)}(x) =  0.5929  \cdot e^{-0.0931 \ x}$. (color online) }
\label{fig:psi}
\end{figure}

The normalized upper and lower components of the Dirac spinor, $\psi_{1,2}(x)$, are plotted in Fig.~\ref{fig:psi}, for the case 
$m=0.1$, $\alpha=1$, $\gamma=0$ and $\lambda=1$. The corresponding probability density $\rho(x)$ is also displayed.
$\psi_1(x)$ is obtained numerically using the shooting method. By inspection of the Eq.~(\ref{Decoupled Dirac Eq}) we see
that the coefficient of $\psi_1'(x)$ is singular when $E+m+\lambda U(x)=0$: this forces the 
first derivative of the wavefunction to vanish at the singularity, represented by a vertical line in the plot. 
Out of this region the wave function
decays exponentially as $\psi_1(x) \propto e^{-\sqrt{m^2-E^2} x}$. The dashed line is a fit of the numerical 
results, within the interval $5 \leq x \leq 50$ and it corresponds to  $\psi_1^{(fit)}(x) =   0.5929 \cdot e^{-0.0931 \ x}$.
Note that $\Gamma^{(fit)} = 0.0931$ is in perfect agreement with the expected expression $\Gamma = \sqrt{m^2-E^2}$.

\begin{figure}[h]
\centering
\includegraphics[width=0.6\textwidth]{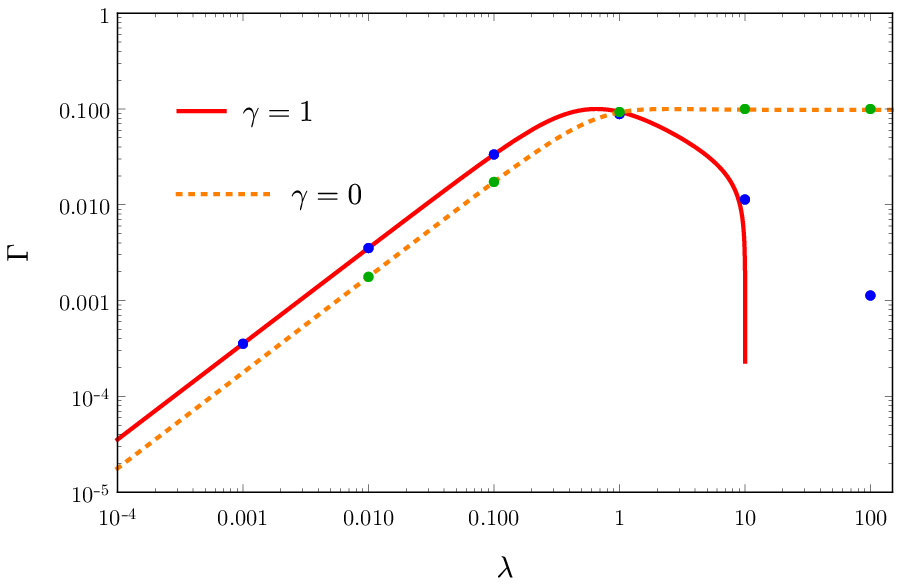} 
\caption{Constant $\Gamma$ of the exponential decay of the wave function. The dots (blue and green) are
the values of $\Gamma$ obtained from the fit of the numerical wave function
at different values of $\lambda$ at $\gamma=1$ and $\gamma=0$ respectively. The solid and dashed 
lines are the explicit expressions obtained using the Pad\'e of Eq.~(\ref{pade}). (color online)}
\label{fig:decay}
\end{figure}

This remarkable agreement can be appreciated from Fig.~\ref{fig:decay}, where the constant $\Gamma$ is extracted
from the fit of the numerical results of the wave function $\psi_1(x)$, at different values of $\lambda$ (the dots
in the plot), and  contrasted with the explicit expressions obtained using the Pad\'e approximant of Eq.~(\ref{pade}).
While, in the non-relativistic case, the wave function decays more and more strongly as $\lambda \rightarrow \infty$,
in the relativistic case the energy of the bound state obeys the inequality $-m < E < m$, and therefore $0< \Gamma \leq m$.
The particular behavior of the analytic formula for $\Gamma$ when $\gamma=0$, which breaks down at $\lambda \approx 10$, is 
easily explained by the fact that the Pad\'e slighlty underestimates the limiting energy for $\lambda > 10$,
 and as a result $\sqrt{m^2-E^2}$ becomes imaginary.

{\sl Conclusions.} 
We have calculated for the first time the energy of a relativistic bound state in a shallow short range potential in one dimension
to fourth order in perturbation theory, proving that the first genuinely relativistic correction appears only at order four.
We have confirmed this generally tiny contribution in a number of cases where it was possible to contrast our results with exact
results available in the literature and with precise numerical calculations, carried out for the case of a pair of gaussian potentials.

We have also shown that it is possible to extend the perturbative analysis to the study of deep wells, by 
using a Pad\'e approximant which captures the asymptotic behavior of the energy for $\lambda \rightarrow \infty$.
The simple analytical formula that we have obtained has been tested for the (not exactly solvable) case of  gaussian
well, finding that the analytical approximation is in excellent agreement with the numerical results.

\section*{Acknowledgements}
The research of P.A. and E.J. was supported by the Sistema Nacional de Investigadores (México).

\end{document}